\documentclass[aps,amsmath,amssymb,groupedaddress,]{article}
\usepackage{authblk}

\usepackage{geometry}
\usepackage{amsmath}
\usepackage{amsfonts}
\usepackage{amsthm}

\usepackage{graphicx}

\usepackage{amssymb}

\usepackage{booktabs}
\usepackage{caption} 
\def\unit#1{\ensuremath{\mathrm{\,#1}}}

\usepackage{xcolor} 

\title{Glass and Jamming Rheology in Soft Particles made of PNIPAM and polyacrylic acid}

\author[1,2]{Silvia Franco}
\author[2]{Elena Buratti}
\author[2,3]{Valentina Nigro}
\author[2,4]{Emanuela Zaccarelli}
\author[2,4]{Barbara Ruzicka}
\author[2,4]{Roberta Angelini}

\affil[1]{Dipartimento di Scienze di Base e Applicate per l'Ingegneria (SBAI), Sapienza Università di Roma, 00185 Roma, Italy;}
\affil[2]{Istituto dei Sistemi Complessi del Consiglio Nazionale delle Ricerche (ISC-CNR), Sede Sapienza, 00185 Roma, Italy;}
\affil[3]{ENEA C.R. Frascati, FSN-TECFIS-MNF Photonics Micro and Nanostructures Laboratory, 00044 Frascati, Rome, Italy;}
\affil[4]{Dipartimento di Fisica, Sapienza Università di Roma, 00185 Roma, Italy;}

\date{March 2021}

\begin{document}

\maketitle

\begin{abstract}

The phase behaviour of soft colloids has attracted great attention due to the large variety of new phenomenologies emerging from their ability to pack  at very high volume fractions. Here we report rheological measurements on interpenetrated polymer network microgels composed of poly(N-isopropylacrylamide) (PNIPAM) and polyacrylic acid (PAAc) at fixed PAAc content as a function of weight concentration.
We found three different rheological regimes characteristic of three different states: a Newtonian shear-thinning fluid, an attractive glass characterized by a yield stress, and a jamming state. We discuss the possible molecular mechanisms driving the formation of these states.
\end{abstract}

\section{Introduction}

In the wide panorama of soft matter, hard spheres have long been the subject of many experimental and theoretical studies \cite{PuseyNature1986, PuseyPRL1987, BradyCOCIS1996, Pham2008, RoyallSM2012}. Similarly, in more recent years, colloids with tunable softness have attracted increasing attention as evidenced by the large number of studies \cite{SenffJCP1999, LikosPhysRep2001, CrassousJCP2008, MattssonNature2009, FernandezNievesWyss, LyonRevPC2012, VlassopoulosCOCIS2014, Bonn2017, NigroSM2017, BergmanNC2018, KargLang2019, GnanNature2019, NinarelloMacromol2019, NigroMacromol2020}.
In fact, thanks to the deformable and compressible particles that can reach non spherical shapes and can be packed in all the available volume, they can give rise to a large variety of new physical states and phenomenologies. 
In particular, at very high volume fractions, colloidal glass transition, that shares similiarities with the glass transition observed upon decreasing temperature in molecular supercooled liquids \cite{PuseyNature1986}, and jamming transition are found.
Although the phenomenology of glass and jamming transitions for soft particle systems are apparently similar \cite{IkedaSM2013, Olsson2007, OHern2003}, many experimetal works \cite{PelletSM2016, ScottiSM2020, GhoshSM2019, BraibantiSciRep2017}, theoretical models and numerical simulations have been proposed \cite{Olsson2007, Tighe2010, Lerner2012, Olsson2012, CharbonneauARCMP2016} to characterize the two phenomena.
In a glass, particles are kinetically trapped in the cage of the first neighbours and their motion occurs in a wider time window than the observation time scale \cite{PuseyNature1986, PuseyJPCM2008}.
When particles are packed at  higher volume fractions, due to their deformability and regardless of thermal fluctuations, they reach the jammed state interpenetrating each other \cite{Tighe2010, ConleyNatComm2019}.
While hard spheres reach the glassy and jammed states for volume fractions $\phi=0.58$ and $\phi=0.64$ respectively, softer particles can be compressed and deformed to reach these states at much  higher volume fractions \cite{PetekidisJPCM2004} whose values are peculiar of the system. 
On the contrary, some studies on PNIPAM microgels have shown that they reach the onset of crystallization at much lower volume fraction compared to hard spheres \cite{KargAFM2011}.
Soft colloidal systems includes emulsions, foams, microgel suspensions and star polymers \cite{Bonn2017, Philippe2018, Dinkgreve2015, MasonBibette1995, ShaoSM2013, Coussot2010}.
Among these, microgels, spherical particles made of crosslinked polymer networks, have emerged as intriguing systems combining the peculiarities of polymers and colloids with distinctive viscoelastic properties and a rich phase behaviour \cite{LyonRevPC2012, KargLang2019, Rovigatti2019, StiegerLang2004}. They can be responsive to external stimuli such as temperature, pH or light \cite{LyonRevPC2012} and this makes microgels widespread in practical applications \cite{KargLang2019, VinogradovCPD2006, ZhouBio2008, BaglioniNanoscale2012, ParkBio2013, MazzucaACSAPM2020}.
In the case of microgels made of the thermoresponsive polymer poly(N-isopropylacrylamide) (PNIPAM), particles undergo the so-called Volume Phase Transition (VPT) at T$\sim$305$\unit{K}$ changing their mutual interactions and their size from a swollen state at low temperature to a compact and shrunken one at high temperature. 
From the molecular point of view, this behaviour can be explained considering that at room temperature PNIPAM polymer is hydrophilic and strongly hydrated in solution, while above 305$\unit{K}$ the affinity to the solvent changes, with PNIPAM becoming less hydrophilic \cite{TavagnaccoMolecular2018}.
Such a Volume Phase Transition is thus an echo at the colloidal scale of the coil-to-globule transition of PNIPAM chains \cite{PeltonColloids1986, FernandezNievesWyss}.
Different states, such as liquid, glass and jammed states have been distinguished by varying control parameters such as packing fraction, crosslinker content and temperature. The liquid to glass to jamming transitions of PNIPAM-based microgel suspensions have been investigated  both theoretically and experimentally in many works \cite{IkedaSM2013, PelletSM2016,  ScottiSM2020, GhoshSM2019, Nordstrom2010,  PaloliSM2013, BasuSoftMatter2014}.
In particular, rheological measurements of PNIPAM microgel suspensions \cite{PelletSM2016, GhoshSM2019} and of concentrated samples of ultra-low crosslinked PNIPAM microgels \cite{ScottiSM2020} reveal that they can form a glass and even achieve a jammed state .
In this work we investigate the rheological behaviour of a microgel composed of Interpenetrated Polymer Networks (IPN) of PNIPAM and polyacrylic acid (PAAc), a thermoresponsive and pH sensitive polymer respectively, at fixed PAAc content (C$_{PAAc}=24.6 \%$) and different weight concentrations.
The great advantage of these systems is the tunability of particles through temperature, pH, PAAc and crosslinker content that provide  variable softness, affect the packing of colloids and hence influence their structural and rheological behaviour.
Therefore, it is challenging to study the dynamics and flow properties of these microgels at molecular level over a wide concentration range.
We found that, with increasing concentration, the system passes from a liquid to a glassy state at the glass transition concentration C$_{w_g}=0.67 \%$ and by further increasing concentration it reaches a jammed state. The behaviour of the storage and loss moduli as a function of shear strain helps to identify the nature of the two states: an attractive glass for $0.6 \% <$ C$_w \leq 3.0 \%$ and a  jammed state for C$_{w_j}>3.0 \%$.
The manuscript is structured as follows: firstly, we report the experimental flow curves of an IPN microgel at C$_{PAAc}=24.6 \%$, at two temperatures below and above the VPT and at different concentrations. They show a viscous behaviour up to C$_{w}=0.6 \%$ and are well fitted through the Cross model, then the emergence of a yield stress makes necessary the introduction of the Herschel-Bulkley model.
Then, the normalized relaxation time, the zero and infinite shear viscosity from the Cross model are compared to the structural relaxation time previously measured through DLS.
Aside the perfect agreement, a  Vogel-Fulcher-Tammann fit allows to determine the glass transition concentration C$_{w_g}=0.67 \%$.
Moreover, for C$_{w}>0.67 \%$ the appearance of a yield stress in the flow curves and its plot versus concentration, allows to determine a glass regime and a jamming transition.
Finally, the investigation of  the storage and loss moduli as a function of frequency and shear strain enables to determine the nature of the two arrested states.

\section{Materials and Methods}

\subsection{Materials}

N-isopropylacrylamide (NIPAM) monomer, acrylic acid (AAc), sodium dodecyl sulphate (SDS) (98$\%$ purity), potassium persulfate (KPS) (98$\%$ purity), ammonium persulfate (APS) (98$\%$ purity), N,N,N’,N’-tetramethylethylenediamine (TEMED) (99$\%$ purity), ethylenediaminetetraacetic acid (EDTA), NaHCO$_{3}$, all from Sigma-Aldrich (St. Louis, Missouri, United States), and the crosslinker N,N’-methylene-bis-acrylamide (BIS), from Eastman Kodak (Kingsport, Tennessee, United States), were used.
NIPAM monomer and BIS, used as crosslinker, were recrystallized from hexane and methanol respectively, dried under reduced pressure (0.01$\unit{mmHg}$) at room temperature and stored at 253$\unit{K}$.
AAc monomer was purified by distillation (40$\unit{mmHg}$, 337$\unit{K}$) under nitrogen atmosphere in presence of hydroquinone and stored at 253$\unit{K}$.
SDS, used as surfactant, KPS and APS, used as initiators, TEMED, a reaction accelerator, EDTA, a chelating agent for purifying dialysis membranes, and NaHCO$_{3}$, were all used as received.
All other solvents were RP grade (Carlo Erba, Cornaredo, Milan, Italy) and were used as received.
Ulatrapure water (resistivity: 18.2$\unit{M} \Omega / cm$ at 298$\unit{K}$) was obtained with Arium\textregistered$\:$ pro Ultrapure water purification Systems, Sartorius Stedim.
A dialysis tubing cellulose membrane, MWCO 14000 Da, (Sigma-Aldrich) was cleaned before use by washing with running distilled water for 3h, treated at 343$\unit{K}$ for 10$\unit{min}$ into a solution containing 3.0$\%$ NaHCO$_{3}$ and 0.4$\%$ EDTA weight concentration, rinsed in distilled water at 343$\unit{K}$ for 10$\unit{min}$ and then in fresh distilled water at room temperature for 2$\unit{h}$.

\subsection{Microgel synthesis}

IPN microgels were obtained by a sequential free radical polymerization method \cite{XiaLangmuir2004, HuAdvMater2004} interpenetrating PAAc network in PNIPAM microgels, synthetized by a precipitation polymerization method following the procedure described by Pelton and al. \cite{PeltonColloids1986}.
In the synthesis, (24.162$\pm$0.001)$\unit{g}$ of NIPAM monomer, (0.4480$\pm$0.0001)$\unit{g}$ of crosslinker BIS and (3.5190$\pm$0.0001)$\unit{g}$ of surfactant SDS  were solubilized in 1560$\unit{mL}$ of ultrapure water and transferred into a 2000$\unit{mL}$ four-necked jacketed reactor equipped with condenser and mechanical stirrer.
Then, the solution, heated at (343$\pm$1)$\unit{K}$, was deoxygenated by purging with nitrogen for 1$\unit{h}$.
The addition of (1.0376$\pm$0.0001) g of KPS, dissolved in 20$\unit{mL}$ of deoxygenated water, allows the polymerization to begin and it is stopped after 5 hours.
The resultant PNIPAM microgel was purified by dialysis against distilled water for two weeks changing water frequently.
Through gravimetric measurements, the final weight concentration C$_{w}$ of PNIPAM micro-particles was determined: C$_{w}$=1.06$\%$.
In the second step of the two-step process for the synthesis of IPN microgels, (140.08$\pm$0.01) g  of the recovered PNIPAM dispersion (C$_{w}$=1.06$\%$) was diluted with ultrapure water up to a volume of 1260$\unit{mL}$ into a 2000 mL four-necked jacketed reactor, kept at (295$\pm$1)$\unit{K}$ by circulating water.
Then, 5$\unit{mL}$ of AAc monomer and (1.1080$\pm$0.0001)$\unit{g}$ of BIS crosslinker were added to the dispersion and the mixture was deoxygenated by bubbling nitrogen inside for 1$\unit{h}$.
0.56$\unit{mL}$ of accelerator TEMED was poured and the polymerization was started with (0.4447$\pm$0.0001)$\unit{g}$ of initiator APS.
The polymerization reaction was stopped by exposing to air and PAAc content depends on the time at which the reaction is stopped.
The resultant IPN was purified by dialysing against distilled water with frequent water changes for two weeks, and then lyophilized and redispersed in water to prepare samples at weight concentration C$_{w}$ = 1.0$\%$.
The PAAc weight concentration (C$_{PAAc}$) of the synthesized IPN sample was determined by both combination of elemental and $^{1}$H-NMR analysis as described in reference \cite{MicaliCPC2018}, and it was C$_{PAAc}$=24.6$\%$.
IPN microgel polydispersity is around 15-20$\%$ and was determined both by DLS and TEM measurements as discussed in reference \cite{NigroJCIS2019}.
Future studies will be devoted to understand the internal strucutre of the IPNs.
Samples at low concentrations (lower than C$_{w}$=1.0$\%$) were obtained by dilution with distilled water from the same stock suspension at C$_{w}$=1.0$\%$ whereas, samples at higher concentrations were obtained by evaporation from the sample C$_{w}$=1.0$\%$ and adjusted at pH=5.5. At this pH a fraction of COOH groups of PAAc chains are protonated while a fraction, not negligible, are deprotonated in the COO$^-$ groups. This affects the aggregation mechanisms that can be ascribed to both like-charge attraction and to intra-particle and inter-particle H-bonding interactions between the CONH group of PNIPAM and the COOH group of PAAc \cite{NigroJML2019}.

\subsection{Rheological measurements}

Rheological measurements have been carried out with a rotational rheometer Anton Paar MCR102 with a cone plate geometry (plate diamater=49.97$\unit{mm}$, cone angle=2.006$^{\circ}$, truncation=212 $\mu$m).
Temperature was controlled using a Peltier system and solvent evaporation was avoided thanks to an isolation hood.
Before measurements, samples were pre-sheared at shear rate $\dot{\gamma}$=500 s$^{-1}$ for 30$\unit{s}$ in order to erase any mechanical history \cite{GhoshSM2019, MenutSM2012} placing them in a reproducible state.
Once loaded, the sample was thermalized for a few minutes before proceeding with the measurement. Measurements on IPN microgels at different weight concentrations in the  range C$_{w}$=(0.05-7.2)$\%$ were performed at T=298$\unit{K}$, when particles are swollen, and at T=311$\unit{K}$, when particles are shrunken. In order to obtain the corresponding packing fractions, the following conversion equation can be used $\zeta=k C_w$ with $k$=233 for C$_{PAAc}=24.6 \%$ \cite{FrancoJPCM2021}.
Two types of rheological tests have been performed: steady shear and oscillatory measurements.

\subsection{Steady shear measurements}
In steady shear tests, the shear stress $\sigma$ was measured as a function of the shear rate $\dot{\gamma}$ varied in the range (10$^{-3}$-10$^{4}$)$\unit{s^{-1}}$ and a stepwise increase of the stress was applied with an equilibration time of 30$\unit{s}$.

\subsubsection{Cross and Carreau-Yasuda models for viscous liquids}
The rheologial behaviour of Newtonian fluids is characterized by the wellknown linear dependence of the shear stress $\sigma$ on  the shear rate $\dot{\gamma}$ and the flow curves are straight lines crossing the origin whose slope is the viscosity. Nevertheless, several phenomenologies emerge in non-Newtonian fluids with a more complex relation between shear stress and shear rate. 
In the case of microgel suspensions at low concentrations, the flow curves at high shear rate resemble those of a shear thinning fluid and data are well described by the Cross model \cite{Macosko, SenffCPS2000, Cross1964}:
\begin{equation}
\sigma(\dot{\gamma})=\dot{\gamma} \left[\eta_{\infty}+\frac{\eta_{0}-\eta_{\infty}}{1+(\tau_C\dot{\gamma})^{m}}\right]
\label{Cmodel}
\end{equation}
where $\eta_{0}$ and $\eta_{\infty}$ are the zero and infinite shear rate limiting viscosities respectively, $m$ a power exponent, $\tau_C$ the relaxation time of the system that marks the onset of shear thinning \cite{PelletSM2016} and its inverse $\dot{\gamma}_{c}$ represents an intermediate critical shear rate.
\\
Another model generally used to reproduce such a phenomenology is the Carreau-Yasuda model that at variance with the Cross model presents an additional fit parameter:
\begin{equation}
\sigma(\dot{\gamma})=\dot{\gamma} \left[\eta_{\infty}+(\eta_{0}-\eta_{\infty}) \left(1+\left(\tau_{CY} \dot{\gamma}\right)^{a}\right)^{b}\right]
\label{CYmodel}
\end{equation}
The two models overlap if $b$=-1, $a$=$m$ and $\tau_{CY}$=$\tau_C$. They can be considered non-Newtonian models since both can be rewritten with the simplified expression $\sigma=\eta(\dot{\gamma})\dot{\gamma}$.
\subsubsection{Yield stress materials: the Herschel-Bulkley model}
With increasing concentration, the flow curves of IPN microgels show a yield stress below which the system does not flow, it represents the stress to apply in order to induce macroscopic flow from rest.
Notoriously, a finite yield stress occurs when, introducing in the system an external deformation, particles motion is so slow that they are no more able to rearrange themselves fast enough to relax the stored stress.
In this case, the structural relaxation time $\tau_{\alpha}$ grows as the viscosity of the system \cite{Bonn2017}.
To describe the flow curves of a yield stress material, the simplest model is the Bingham one \cite{Bingham} according to which the system needs an initial yield stress to start to flow but then, the shear stress increases steadily with increasing shear rate as in Newtonian fluids.
Therefore, this model requires two parameters to be described,  the yield stress $\sigma_{y}$ and the so called plastic viscosity $\eta_{p}$  representing the slope of the flow curve in the fluid region:
\begin{equation}
\dot{\gamma}=0 \; \; \; \; \; \; \; \; \; \; \; \; \; \; \; \; \;  \sigma<\sigma_{y}
\label{Bingham1}
\end{equation}
\begin{equation}
\sigma=\sigma_{y}+\eta_{p} \dot{\gamma}  \: \: \; \; \; \;   \sigma<\sigma_{y}.
\label{Bingham2}
\end{equation}
\\
However, concentrated colloidal systems generally behave in a more complex way and rarely follow the Bingham model that is a specific case of the more popular and widely used Herschel-Bulkley (HB) model \cite{Barnes} that takes into account also the non-Newtonian nature of the fluid after the plastic rearrangements induced by the yield stress:
\begin{equation}
\sigma(\dot{\gamma})=\sigma_{y}+k \dot{\gamma}^u
\label{HBmodel}
\end{equation} 
where $\sigma_{y}$ is the yield stress, for $\sigma<\sigma_{y}$  there is no flow and the system behaves as a solid, for $\sigma>\sigma_{y}$ instead it flows, $k$ is named ``consistency'' index 
and $u$ is the flow index that defines the non-Newtonian behaviour.
Values $u<$1 are typical of shear thinning fluids and $u>$1 defines systems characterized by shear thickening \cite{Bonn2017, GhoshSM2019, Mueller2010}, whereas if $u$=1 the Bingham model is recovered.
The presence of a finite yield stress arises as the glass or jamming transitions are crossed \cite{VlassopoulosCOCIS2014, Bonn2017, GhoshSM2019, NakaishiMacromol2018}. 
\\
When the dependence of $\sigma_{y}$ on concentration C$_{w}$ is strong, the system is supposed to be in the ``glassy regime".
Whereas, when the increase is rather linear, the system is probably in the ``jamming regime" that can be reached at higher concentrations \cite{IkedaSM2013, GhoshSM2019}.
Moreover, passing from glassy to jamming regime the flow index $u$ decreases, with increasing concentration, up to values less than 0.45 approaching the jammed state \cite{ IkedaSM2013, Olsson2007, OHern2003, PelletSM2016, ScottiSM2020, GhoshSM2019, Nordstrom2010}.
\\
The Herschel-Bulkley model can be also written in a modified form \cite{Bonn2017, GhoshSM2019}:
\begin{equation}
\sigma(\dot{\gamma})=\sigma_{y} \left( 1+ \left( \frac{\dot{\gamma}}{\dot{\gamma}_{c}} \right)^{n} \right)
\label{HBmodel2}
\end{equation}
where the characteristic shear rate $\dot{\gamma}_{c}$:
\begin{equation}
\dot{\gamma}_{c}=\left( \frac{\sigma_{y}}{k} \right)^{1/n}
\label{gammac}
\end{equation}
determines the transition from a rate independent plastic flow to a rate dependent viscous flow \cite{GhoshSM2019}. 
Also the  Herschel-Bulkley model is a  generalized Newtonian model. 

\subsection{Oscillatory measurements: Storage and Loss moduli}
The rheological dynamic characterization of microgels has been carried out also in sinusoidal regime allowing to extend all rheological quantities to the frequency domain \cite{FernandezNievesWyss}.
The experimental procedure requires the application of an oscillatory tangential deformation, shear strain $\gamma$(t):
\begin{equation}
\gamma(t)=\gamma_{0}sin(\omega t)
\end{equation}
where $\omega$=2$\pi$f is the angular frequency, being f the frequency and $\gamma_{0}$ is a fixed amplitude.
If the studied material presents a linear viscoelastic behaviour, it will respond with a sinusoidal shear stress shifted by an angle $\delta$ with respect to the applied deformation whose value will depend on the characteristics of the material:

\begin{equation}
\sigma(t)=\sigma_{0}sin(\omega t+\delta)
\end{equation}
It can be rewritten as:
\begin{equation}
\sigma(t)=\gamma_{0}G'(\omega)sin(\omega t)+ \gamma_{0}G''(\omega)cos(\omega t)
\end{equation}
where G$'$($\omega$) is the \textit{storage modulus}, in phase with deformation $\gamma$, and  G$''$($\omega$) is the \textit{loss modulus}, in opposite phase with deformation.
They represent, respectively, the fraction of elastic energy stored and subsequently released by the system per cycle per unit volume of deformation, and the energy dissipated as heat per cycle by viscous effects.
When $\mathrm{G}' \gg \mathrm{G}'' $ 
and $\delta=0^{\circ}$ the system behaves like a solid and when G$'<''$ and $\delta$=90$^{\circ}$ like a liquid. The system is vicoleastic when G$'\approx$G$''$ or 0$^{\circ}<\delta<90^{\circ}$ \cite{FernandezNievesPuertas}.
The study of G$'$ and G$''$ provides numerous interesting information on dynamical and phase behaviour of the microgels through their phase transitions \cite{FernandezNievesWyss}. 
Here we investigated moduli as a function of frequency (\textit{frequency sweeps}) and shear strain (\textit{amplitude or strain sweeps}).
\\
Frequency sweep tests were performed in the frequency range $f = (0.01-100)\unit{Hz}$ at fixed strain $\gamma$ to probe the linear response of the samples.
\\
Frequency sweeps were performed in \textit{linear viscoelastic regime} in fact when the applied strain $\gamma$ is small enough, most of viscoelastic materials behave linearly with direct proportionality between the stress $\sigma$ and the deformation $\gamma$ \cite{Rubinstein}. In this regime the moduli are independent of applied strain.
In order to ensure that frequency sweep measurements were carried out within linear viscoelastic regime, before any frequency sweep, a strain sweep test was performed  at fixed frequency 1$\unit{Hz}$ to determine the most suitable strain value for each sample chosen within the initial constant region.
Amplitude or strain sweeps were carried out in the strain range $\gamma$=(0.1-10)$\%$  at fixed frequency. Increasing the strain amplitude, two regimes are identified \cite{CarmonaJFE2014}. In the first one, at small amplitude oscillatory shear (SAOS), the material shows a linear viscoelastic response, in the second one, at large amplitude oscillatory shear (LAOS), a non linear viscoelastic response occurs.
\\
\subsection{Arrhenius and Vogel-Fulcher-Tamman models for the concentration dependence of viscosity and relaxation time}

Viscosity of liquids approaching the glass transition depends strongly on concentration and/or temperature and it usually increases with increasing concentration or decreasing temperature, this behaviour microscopically corresponding to a slowing down of the dynamics. When viscosity $\eta$ shows a low sensitivity to small changes of the control parameters (C$_w$, T...) it follows the well known exponential Arrhenius-like behaviour \cite{ChemicalKinetics} that in the case of concentration can be 
written as:
\begin{equation}
\eta=\eta_{0} \, exp \Big( A C_w \Big)
\label{Arrhenius}
\end{equation}
where $\eta_{0}$ is the viscosity in the limit of C$_w$=0 and A controls the growth of the function.
Conversely, when $\eta$ shows a very steep increase with respect to small changes in concentration, viscosity is well described by the Vogel-Fulcher-Tammann (VFT) model \cite{Angell2000} identified by an exponential with three free parameters:
\begin{equation}
\eta=\eta_{0} \, exp \Big( \frac{A C_w}{C_{w_{0}}-C_w} \Big)
\label{VFT}
\end{equation}
where $\eta_{0}$ is the viscosity in the limit of C$_w$=0, A is the growth parameter and C$_{w_{0}}$ is the critical concentration that signs the divergence of $\eta$.
Both behaviours have been observed in a wide range of systems \cite{Angell2000, LuSMS2013}
including soft materials \cite{NigroSM2017, NigroMacromol2020, FrancoJPCM2021, BerthierBiroli2011, SenguptaJCP2011, HunterWeeks2012, VanDerScheerACSN2017} and are precursor of a glass transition that happens at the glass transition temperature T$_{g}$ or concentration C$_{w_{g}}$  .
In the renowned Angell classification \cite{Angell2000, VanDerScheerACSN2017}, by cooling the system up to T$_{g}$ the typical system time scale becomes of the order of 10$^{2}$ $\unit{s}$ and it is accompanied by a simultaneous increase of the viscosity that, around the glass transition, is of the order of 10$^{12}$ \unit{Pa \cdot s} \cite{Angell2000}.
Systems that follow the Arrhenius law are denoted as ``strong glasses'' and those that follow the VFT model are defined as ``fragile glasses'' \cite{AngellScience1995}.
Soft and compressible particles, such as microgels, undergo the glass transition when their concentration approaches the critical value. These systems, depending on softness, can show an Arrhenius or a Vogel-Fulcher-Tamman behaviour resembling that of molecular glasses \cite{NigroSM2017, NigroMacromol2020, FrancoJPCM2021, VanDerScheerACSN2017}.

\subsection{Characteristic stress and time scales}
The microscopic behaviour of suspensions of soft particles  can be described considering the thermal energy scale, $k_{B}T$, the particle interaction energy scale, $\varepsilon$, and their ratio $k_{B}T/\varepsilon$ that is often used as a control parameter \cite{IkedaSM2013, PelletSM2016, IkedaPRL2012}.
Typically, for soft microgels $k_{B}T/\varepsilon$ $\sim$ 10$^{-4}$ whereas for hard spheres  $k_{B}T/\varepsilon$ $\sim$ 10$^{-8}$ \cite{Bonn2017, IkedaSM2013}, this quantity is considered an indicator of particle softness.
At fixed $\varepsilon$, the glass transition can be studied  in the limit of temperature $T\rightarrow0$.
Whereas, keeping the temperature constant in the limit $\varepsilon\rightarrow\infty$, the athermal jamming transition can be investigated.
The two energy scales $k_{B}T$ and $\varepsilon$ are associated to characteristic time and stress scales \cite{IkedaSM2013} that, for a system of particles moving by Brownian motion in dilute conditions and  perturbated by thermal fluctuations, can be expressed respectively as \cite{IkedaSM2013, Philippe2018}:
\begin{equation}
\tau_{T}=\frac{6\pi \eta R^{3}}{k_{B}T}=\frac{R^2}{D_{0}}
\label{eq1}
\end{equation}
\begin{equation}
\sigma_{T}=\frac{k_{B}T}{R^{3}}
\label{eq2}
\end{equation}
where the subscript \textit{T} stands for ``thermal", $k_{B}$ is the Boltzmann constant, $\eta$ represents the solvent viscosity at temperature T, R is the particle radius and $D_{0}$ the self-diffusion coefficient in dilute conditions.
The time scale (equation (\ref{eq1})) is associated with the self diffusion of individual particles namely the time needed to a particle to cover a distance equal to its radius, the quantity $\dot{\gamma} R_{h}^{2} /D_{0}$ is the Peclet number.
The stress scale (equation (\ref{eq2})) is the typical stress created by thermal fluctuations \cite{VlassopoulosCOCIS2014, Bonn2017}.
equation \ref{eq1} and equation \ref{eq2} are often used in rheology to normalize shear rate $\dot{\gamma}$ and shear stress $\sigma$. 

\subsection{Dynamic Light Scattering measurements}

\begin{figure}[h!]
\centering
\includegraphics[trim=0.1cm  4  0.1cm  1, clip,  scale=0.34]{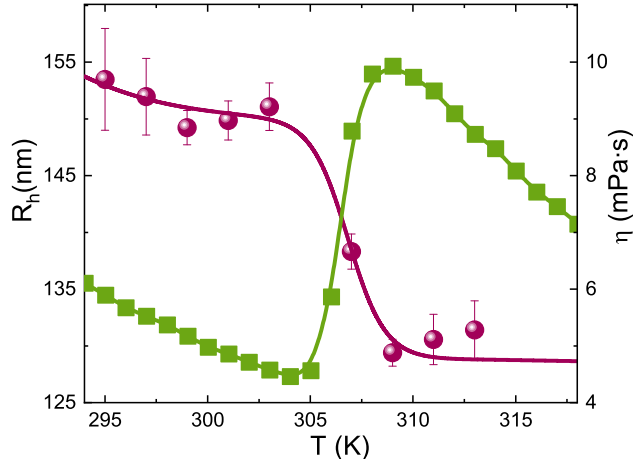}
\caption{Hydrodynamic radius R$_{h}$ (left axis, circles) for IPN microgel at C$_{PAAc}=24.6 \%$ as a function of temperature measured in diluite conditions (C$_{w}$=0.01$\%$) compared with the viscosity (right axis, squares) at concentration C$_{w}$=0.3$\%$ and shear rate 10 s$^{-1}$.}
\label{fig1}
\end{figure}

Particle size of IPN microgels has been characterized through Dynamic Light Scattering (DLS) measurements that provided the hydrodynamic radius, R$_{h}$, as a function of temperature T.
An optical setup based on a solid state laser (100$\unit{mW}$) with monochromatic wavelenght $\lambda$= 642$\unit{nm}$ and polarized beam has been used to probe IPN microgels suspensions in dilute regime.
Measurements have been performed at a scattering angle $\theta$=90$^{\circ}$ that corresponds to a scattering vector $Q$=(4$\pi$n/$\lambda$) sin($\theta$/2)=0.018$\unit{nm}^{-1}$. 
The hydrodynamic radii have been obtained  through the Stokes-Einstein relation: R$_{h}=k_{B}T/6 \pi \eta_{s} D_{t}$ where $k_{B}$ is the Boltzmann constant, $\eta_{s}$ the viscosity of the solvent, namely water, at the measured temperature and $D_{t}$ the translational diffusion coefficient related to the relaxation time $\tau$ through the relation: $\tau=1/(Q^{2} D_{t})$.
The relaxation time was obtained by fitting the autocorrelation function of scattered intensity through the Kohlrausch-William-Watts expression \cite{Kohlrausch1854, Williams1970}, $g_{2}(Q,t)$ = $1+b[exp (-(t / \tau)^{\beta})]^{2}$, with the stretching exponent $\beta$ providing the deviation from the single exponential.
Measurements have been performed as a function of temperature for IPN microgels at C$_{PAAc}$=24.6 \% and are reported in Figure \ref{fig1}.

\section{Results and Discussion}

Figure 1 shows the viscosity $\eta$ versus temperature for an IPN microgel at C$_{PAAc}=24.6\%$, C$_{w}$=0.3$\%$ and  $\dot{\gamma}$= 10$\unit{s^{-1}}$  compared to the hydrodynamic radius R$_{h}$, measured by DLS in very diluite conditions (C$_{w}$=0.01$\%$).
%
%
At low temperature, $\eta(\dot{\gamma})$ displays a slow decrease until a sharp growth takes place followed by a decrease further increasing T.
This behaviour, in agreement with viscosity measurements reported in references \cite{HuAdvMater2004, XiaJCR2005}, emerges in correspondence of the volume phase transition, when particles collapse from a swollen to a shrunken state, as evidenced by the sudden drop of the particle radius.
The viscosity increase on heating is a counterintuitive phenomenon observed previously in few other systems \cite{AngeliniPhilMag2007, AngeliniPRE2008}. In the case of IPN microgels it can be explained considering that, with increasing temperature above the VPT, the collapse of PNIPAM networks  enhances the exposure of PAAc dangling chains and the formation of aggregates providing an abrupt increase of viscosity also favored by a shear-induced association of microgel particles via inter-chain interactions \cite{HoweACIS2009}. The following decrease of viscosity at high temperature can be interpreted as a fluidification of the system after aggregation.

\begin{figure}[h!]
\centering
\includegraphics[trim=0.3cm  0.05cm  0.05cm  0.01cm, clip,  scale=0.32]{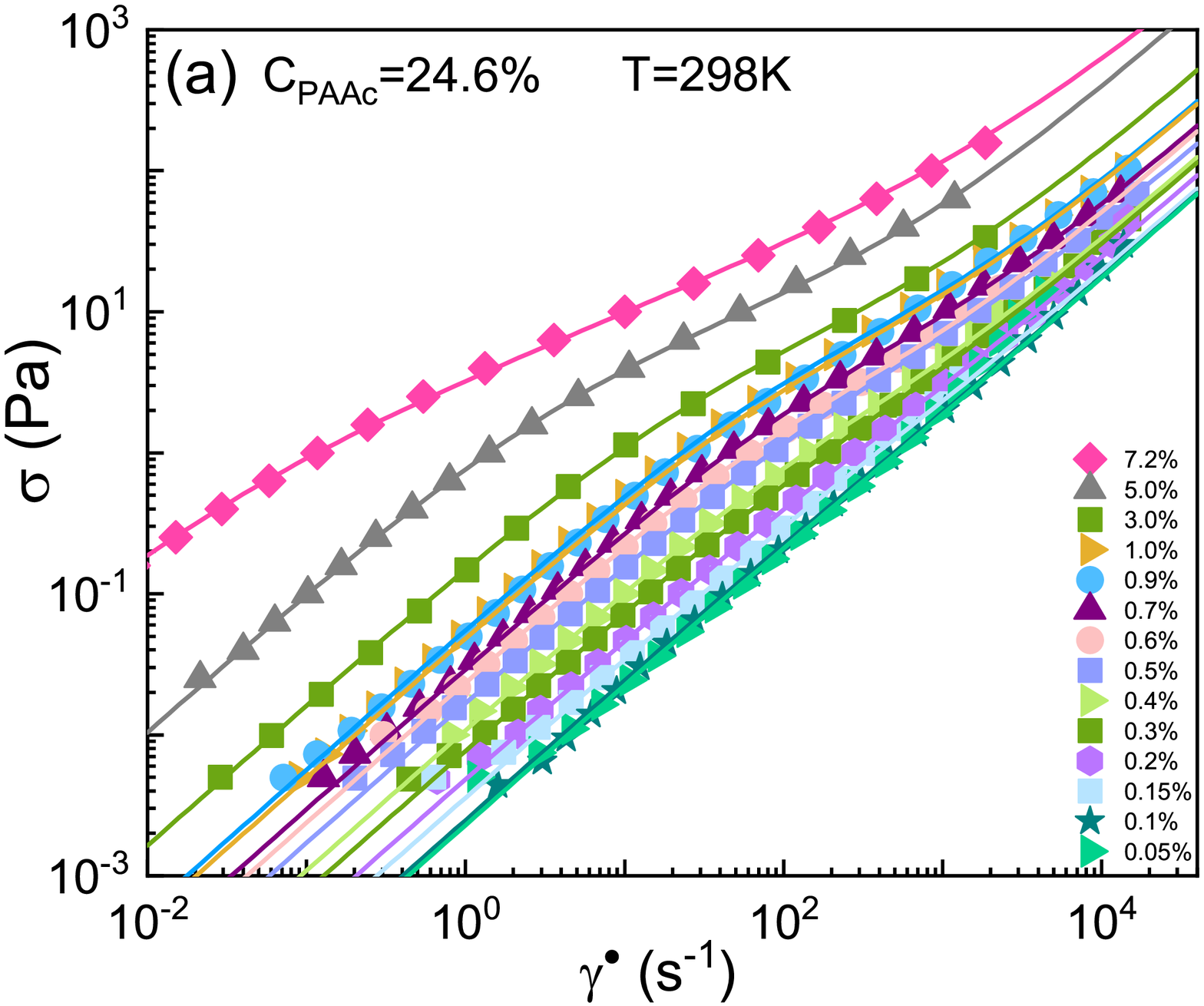}
\quad \includegraphics[trim=0.3cm  0.05cm  0.05cm  0.2, clip,  scale=0.32]{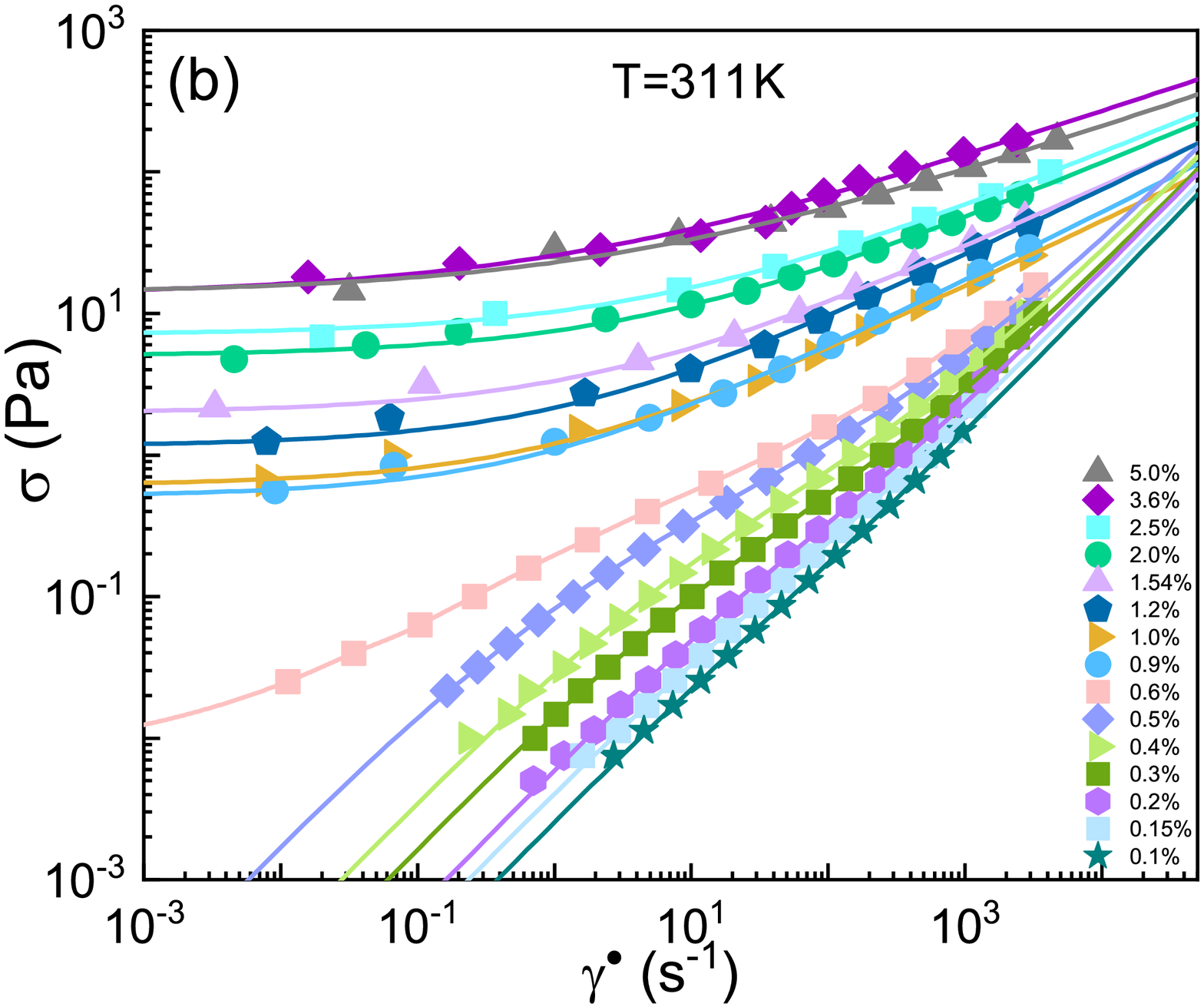}
\caption{Shear stress $\sigma$ versus shear rate $\dot{\gamma}$ for IPN microgel with PAAc content 24.6$\%$ at different weight concentrations in the range (0.05-7.2)$\%$ at \textbf{(a)} T=298$\unit{K}$, below the VPT, and \textbf{(b)} T=311$\unit{K}$, above the VPT. Lines are fits with the Cross model for low concentrations and with the Herschel-Bulkley model for concentrations showing a yield stress.}
\label{fig2}
\end{figure}

In Figure \ref{fig2} the shear stress $\sigma$ as a function of shear rate $\dot{\gamma}$ at T=298$\unit{K}$ in the weight concentration range (0.15-7.2)$\%$ and at T=311$\unit{K}$ in the weight concentration range (0.1-5.0)$\%$ is reported.
%
%
%
%
At T=298$\unit{K}$, below the VPT, the flow curves show a viscous behaviour, in particular at low concentrations they show an almost Newtonian behaviour with  $\sigma \propto \dot{\gamma}$ while at increasing C$_w$ they start to deviate toward a shear thinning dependence.
All data are well described by the Cross model (equation \ref{Cmodel}) all over the investigated concentration range, the fit parameters are reported in Figure S1 of the Supplementary Information (SI).
Analogous results are obtained by using the  Carreau-Yasuda model (equation \ref{CYmodel}) as shown in Figure S2 of the SI where a comparison between the two models is reported with all the fitting parameters. 
Conversely, at T=311$\unit{K}$, above the VPT, data show the same phenomenology observed at T=298$\unit{K}$ only up to C$_{w}$=0.6$\%$ and are well described by the Cross model, while for C$_{w} \geq 0.7\%$ a yield stress appears and data follow the Herschel-Bulkley law (equation \ref{HBmodel}) (all the fitting parameters are reported in Figure S3 of the SI).
The emergence of a finite yield stress $\sigma_{y}$, with increasing weight concentration, marks the transition from a fluid to an arrested state \cite{Bonn2017, PetekidisJPCM2004, Philippe2018}  where the dynamics are slower.
In particular, suspensions which are glassy at rest, show a stress which becomes independent of $\dot{\gamma}$ as $\dot{\gamma} \rightarrow 0$ \cite{PetekidisJPCM2004}.
This behaviour is caused by molecular interactions among microgels due to the exposure of PAAc dangling chains resulting from the collapse of PNIPAM at high temperature. A different phenomenology is observed at T=298$\unit{K}$, in the same concentration range, where PNIPAM networks are swollen inside the particles and thus the interactions due to the PAAc network are reduced.
As a consequence, a higher particle concentration is necessary in order to reach the same aggregation state.
Finally, we observe that for C$_{w}\geq 3.6 \%$ shear stress curves
do not show any further evolution on increasing concentration. 
These results are in perfect agreement with what previously reported for the same system in reference \cite{NigroMacromol2020}
where the microscopic dynamics investigated throught DLS and X-ray photon correlation spectroscopy showed a clear transition from a fluid to a glassy state at almost the same concentration, as evidenced by the behaviour of the normalized intensity autocorrelation functions $g^{2}(q,t)-1$ and of the relaxation time at T= 311$\unit{K}$.
%
%

\begin{figure}[h!]
\centering
\includegraphics[trim=0.3cm  0.05cm  0.05cm  0.01cm, clip,  scale=0.35]{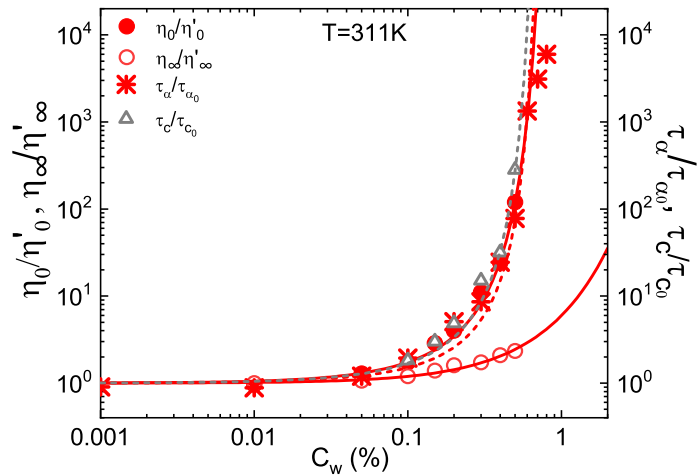}
\caption{Normalized parameters $\eta_0$, $\eta_{\infty}$ and $\tau_{C}$ from the Cross fits to the flow curves of Figure 2a, compared with the structural relaxation time $\tau_{\alpha}$ measured trough DLS for IPN microgel with PAAc content 24.6$\%$ at T=311$\unit{K}$, above the VPT. Lines are fits with the Arrhenius-like model for $\eta_{\infty}$ and with the VFT model for $\eta_{0}$, $\tau_{\alpha}$ and $\tau_{C}$.}
\label{fig3}
\end{figure}

With the aim to higlight the excellent agreement, Figure \ref{fig3} reports the comparison among the normalized relaxation time  $\tau_{C}$, the zero and infinite shear viscosity from the Cross model and the structural relaxation time $\tau_{\alpha}$ from DLS as a function of concentration up to C$_{w}=0.6 \%$ at T=311$\unit{K}$.
%
%
Data grow steeply according to Vogel-Fulcher-Tammann behaviour (equation  \ref{VFT}) with C$_{w_{0}}=0.8\%$, while the glass transition concentration C$_{w_{g}}$, obtained as the C$_{w}$ at which relaxation time is 100 s, is C$_{w_{g}}=0.67 \%$, value in perfect agreement with the appearance of a yield stress in Figure \ref{fig2}(b). This indicates that at high temperature, when particle are collapsed, IPN microgels have a fragile-like  behaviour \cite{NigroSM2017,FrancoJPCM2021} resembling that of hard colloids \cite{Philippe2018} and other microgels \cite{RomeoJCP2012}.
For these systems, C$_{w_{g}}$ corresponds to a generalized volume fraction $\zeta_g$=1.56 \cite{FrancoJPCM2021}, value much higher than the hard sphere volume fraction $\phi_{g}=0.58$ at the glass transition. This can be explained considering that the softness of microgel particles allows them to be compressed
well above 1 since they can balance the increase of osmotic pressure by deforming and adjusting their volume.
%
%
%
%

\begin{figure}[t]
\centering
\includegraphics[trim=0.2cm  2  0.8  0.2cm, clip,  scale=0.32]{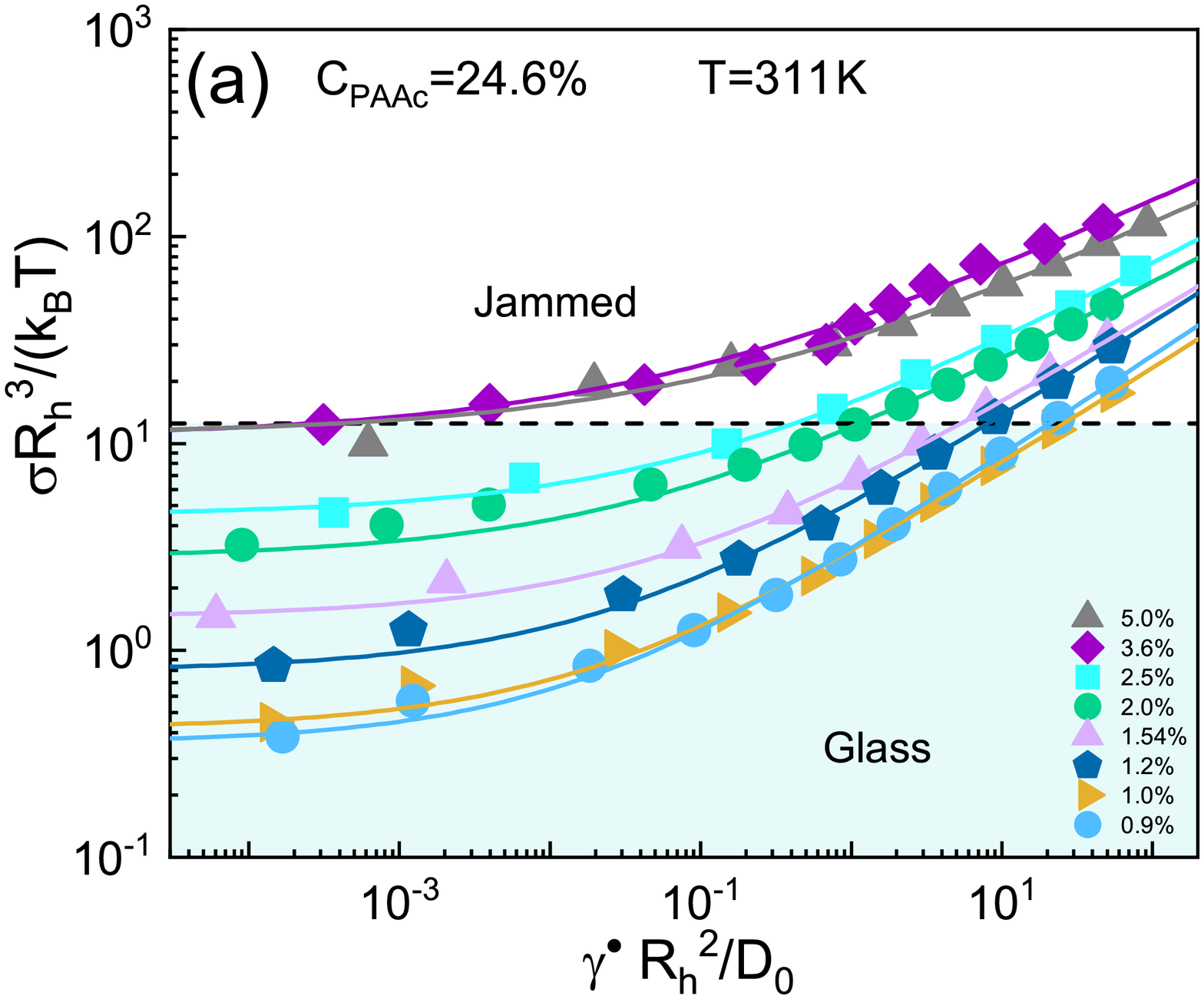}
\quad \includegraphics[trim=0.3cm  2  0.8  0.01cm, clip,  scale=0.32]{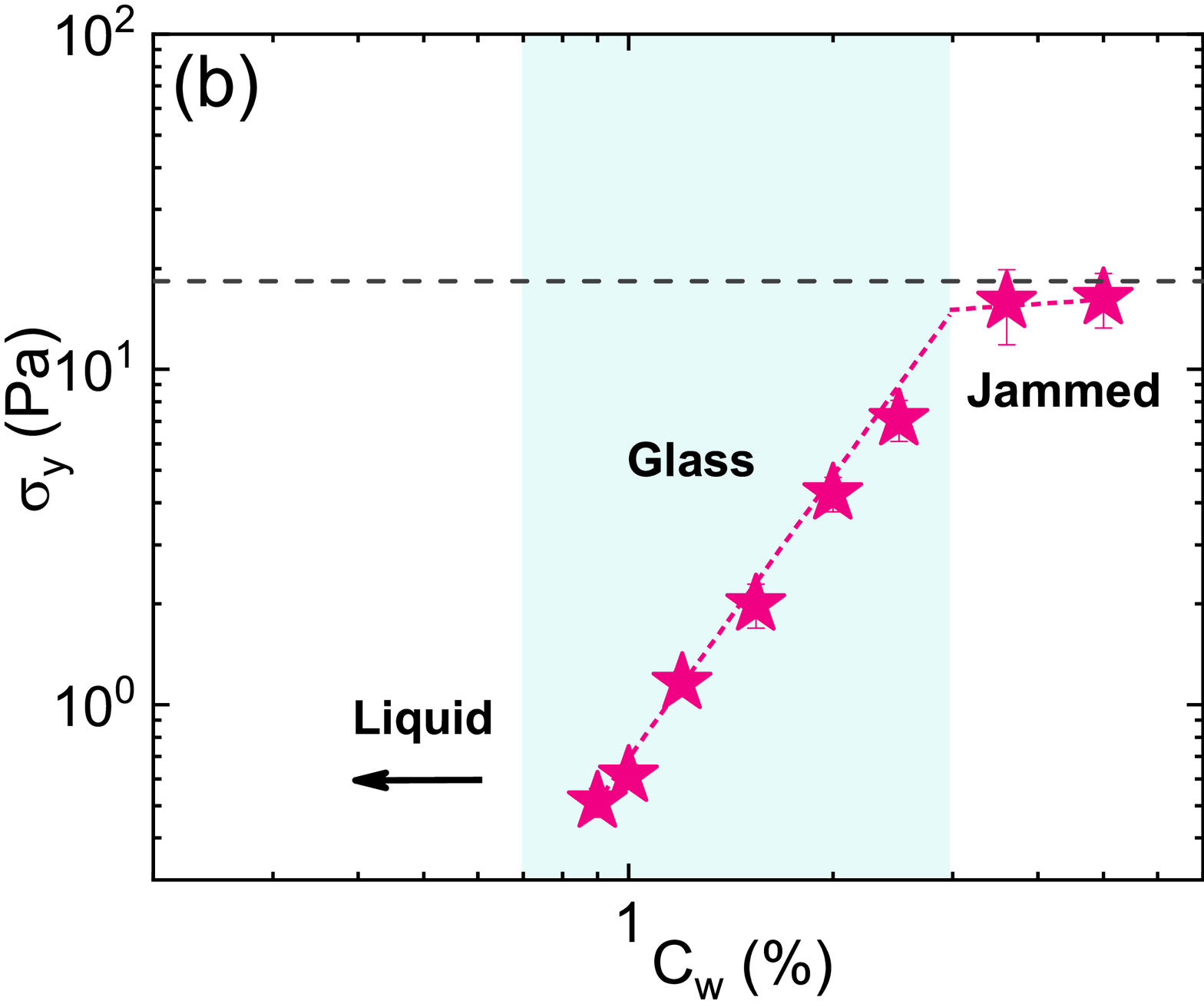}
\caption{\textbf{(a)} Normalized stress versus Peclet number. Lines are fits following the Herschel-Bulkley model.
 \textbf{(b)} Apparent yield stress (pink stars), obtained from the Herschel-Bulkley fit to data in (a), versus weight concentration for IPN with C$_{PAAc}=24.6 \%$ at T=311$\unit{K}$.
The black dashed lines represent the value $\sigma_{y}$R$^{3}$/(k$_{B}$T)=12.5 it is considered the value that nearly marks the border between glass and jammed state as described in the text.}
\label{fig4}
\end{figure}

To better understand the nature of the arrested state, the flow curves  presenting a yield stress $\sigma_{y}$ in Figure \ref{fig2}(b), are plotted  in dimensionless units in Figure \ref{fig4}(a).
Shear stress is normalized according to equation \ref{eq2} dividing by (k$_{B}$T)/R$_{h}^{3}$ and shear rate, normalized according to equation \ref{eq1} multiplying by R$_{h}^{2}$/D$_{0}$, is expressed in terms of Peclet number.
Following the classification of many works \cite{IkedaSM2013, PelletSM2016, ScottiSM2020}, the system can be considered in the glassy regime when $\sigma_{y}$R$^{3}$/(k$_{B}$T)$<$12.5. We can therefore conclude that IPN microgels with C$_{PAAc}$=24.6$\%$ are in the glassy state for $0.7\% \leq$ C$_w$ $\leq$ 2.5 $\%$.
The same classification defines a system in the jamming regime when $\sigma_{y}$R$^{3}$/(k$_{B}$T)$>$12.5 a condition that in our system occurs at C$_w \geq 3.6\%$.
To evaluate the yield stress dependence on concentration, in Figure \ref{fig4}(b) $\sigma_{y}$ obtained from Herschel-Bulkley fits (equation \ref{HBmodel})  is plotted as a function of C$_{w}$.
At low concentrations, where no yield stress occurs ($\sigma_{y}$=0), the system can be considered a viscous liquid.
On the contrary, for $0.7\% \leq$ C$_w \leq 2.5 \%$ a steep increase of the yield stress takes place and a strong concentration dependence ($\sigma_{y} \sim$ C$_{w}^{2.72}$) is shown indicating that the system is in the glassy state \cite{PelletSM2016, ScottiSM2020, GhoshSM2019}.
Further increasing C$_{w}$, the yield stress has a slower increase growing almost linearly with concentration and this can be referred to as the jamming regime \cite{PelletSM2016, GhoshSM2019}. From the intersection of the two fits in Figure \ref{fig4}(b) we can determine the jamming transition concentration C$_{w_j}\approx 3.0\%$.
In this state, particles are in direct mechanical contact with each other and share many similarities with granular materials at jamming \cite{CharbonneauARCMP2016}.
%
%

\begin{figure}[h!]
\centering
\includegraphics[trim=0.2cm  3  1  0.05cm, clip,  scale=0.38]{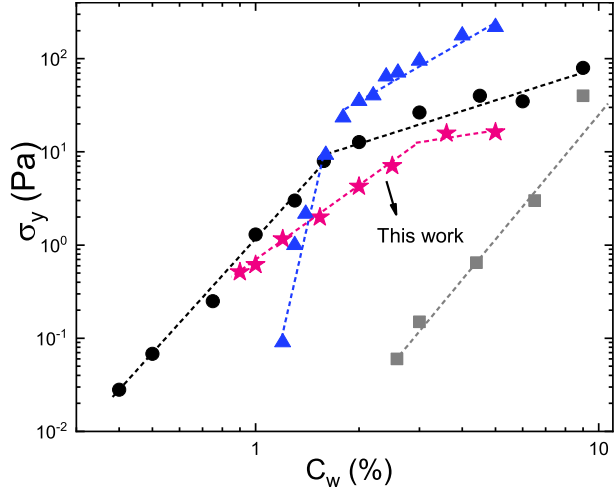}
\caption{Yield stress for microgel with C$_{PAAc}= 24.6 \%$ at T=311$\unit{K}$ (pink stars) obtained from the Herschel-Bulkley fit to data of Figure 2(b) compared with data on PNIPAM microgel (black circles) by Ghosh and coworkers \cite{GhoshSM2019}, on soft polyelectrolyte microgel suspensions (blue traingles) by Pellet and Cloitre \cite{PelletSM2016} and on ultralow crosslinked PNIPAM microgel (grey squares) by Scotti and coworkers \cite{ScottiSM2020}.}
\label{fig5}
\end{figure}

Figure \ref{fig5} reports a comparison of yield stress versus concentration with other experimental studies on microgels \cite{GhoshSM2019, PelletSM2016, ScottiSM2020}.
They are all characterized by an initial sharp growth in the glassy state followed by a smoother growth in the jamming regime.
In particular, for ionic microgel suspensions studied by Pellet and Cloitre \cite{PelletSM2016}, $\sigma_y$ grows sharply with concentration in the glassy regime and displays also a strong concentration dependence in the soft jamming regime.
Instead, in PNIPAM suspensions investigated by Ghosh and coworkers \cite{GhoshSM2019} the increase of yield stress is less pronounced and very similar to that ultralow crosslinked PNIPAM microgel investigated by Scotti and coworkers   \cite{ScottiSM2020}, it is then followed by a nearly constant value in the soft jamming regime. This behaviour is more similar to that of our IPN microgels.
%
%

\begin{figure}[h]
\centering
\includegraphics[trim=0.1cm  2  1  0.2cm, clip,  scale=0.38]{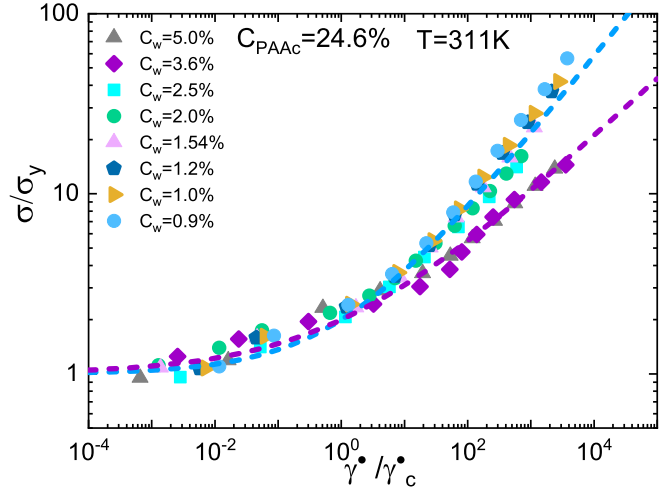}
\caption{Scaled plot of normalized shear stress against reduced shear stress for IPN microgels at C$_{PAAc}$=24.6$\%$, T=311$\unit{K}$ and at different weight concentrations. All data follow two mastercurves indicating the existence of glass and jammed states. Lines are fits to data according to equation \ref{HBmodel2}.}
\label{fig6}
\end{figure}

To better visualize the existence of two different arrested states, the shear stress normalized to the yield stress $\sigma/ \sigma_{y}$ is reported in Figure \ref{fig6} as a function of $\dot{\gamma}$/$\dot{\gamma_{c}}$ where $\dot{\gamma_{c}}$ is the characteristic shear rate of equation \ref{gammac}.
Data at all weight concentrations C$_{w}$ fall into two different mastercurves, one for the glassy behaviour at $0.7 \% \leq$ C$_{w} \leq 2.5 \%$ and a second one for the jamming regime at $3.6 \% \leq$ C$_w \leq 5.0 \%$, confirming the distinction between two different regimes reported in Figure 4.
To corroborate the nature of the different investigated states, we measured the frequency dependence of the storage (G$'$) and loss (G$''$) moduli in the SAOS regime reported in Figure \ref{fig7} as a function of frequency at three different concentrations \cite{LiuPolymers2012} in the liquid, glassy and jamming regimes.
At C$_{w}$=0.4$\%$ the growing trend of the moduli is characteristic of the liquid state.
In the intermediate concentration C$_{w}$=0.9$\%$, G$'$ is greater than G$''$ over the entire frequency range, indicating a viscoelastic solidlike behaviour typical of  glasses \cite{MasonWeitzbis1995, KoumakisSM2008, Carrier2009}.
The loss modulus shows a slight minimum as also observed in other soft glassy materials \cite{MasonBibette1995, MasonWeitz1995, MenutSM2012, Derec2003}, this feature could point out two different processes \cite{MenutSM2012}: at low $f$, a relaxation process resulting from very slow structural time rearrangements, related to the transition from in-cage particle diffusion to a slower, long distance, out-of-cage motion ($\beta$ or fast or microscopic relaxation) \cite{Pham2008, KoumakisSM2008}; at high $f$, the rise of G$''$.
These phenomena reflect the viscous dissipation of the solvent \cite{Carrier2009}.
Briefly, for this concentration the plateau modulus is not perfectly constant but increases with frequency showing weak solid-like properties and no crossover between G$'$ and G$''$ occurs in the probed frequency range, indicating that the system does not manifest significant diffusion or structural relaxation in these time scales.
At C$_{w}$=3.6$\%$, moduli further increase and G$'$ is always greater than G$''$ by over an order of magnitude and is frequency independent revealing a solid-like response of the system.
These results are in good agreement with the finding reported in reference \cite{ZhouBio2008}.

\begin{figure}[h]
\centering
\includegraphics[trim=0.1cm  2  1  0.05cm, clip,  scale=0.42]{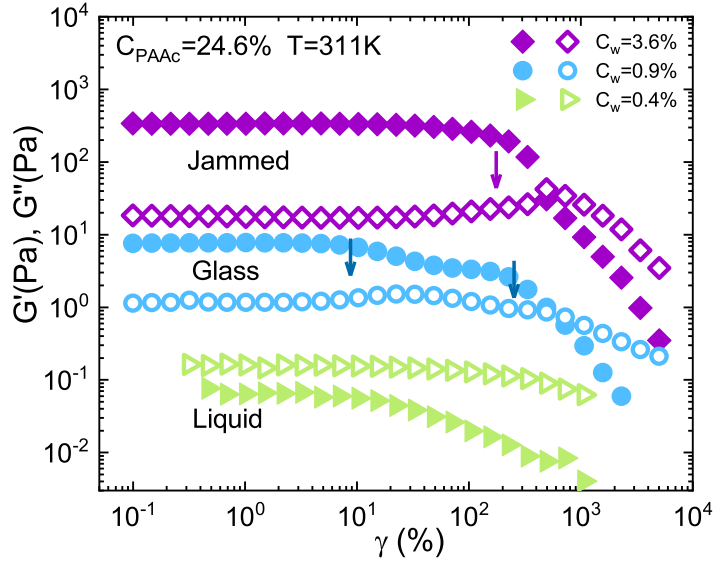}
\caption{Storage G$'$ (filled symbols) and Loss G$''$ (open symbols) moduli versus strain $\gamma$ for IPN microgels at C$_{PAAc}$=24.6$\%$, T=311$\unit{K}$ and at different weight concentrations in the glass and jammed states.
The arrows indicate the different breaking points.}
\label{fig8}
\end{figure}

\begin{figure}[h!]
\centering
\includegraphics[trim=0.22cm  3  0.2  0.1cm, clip,  scale=0.42]{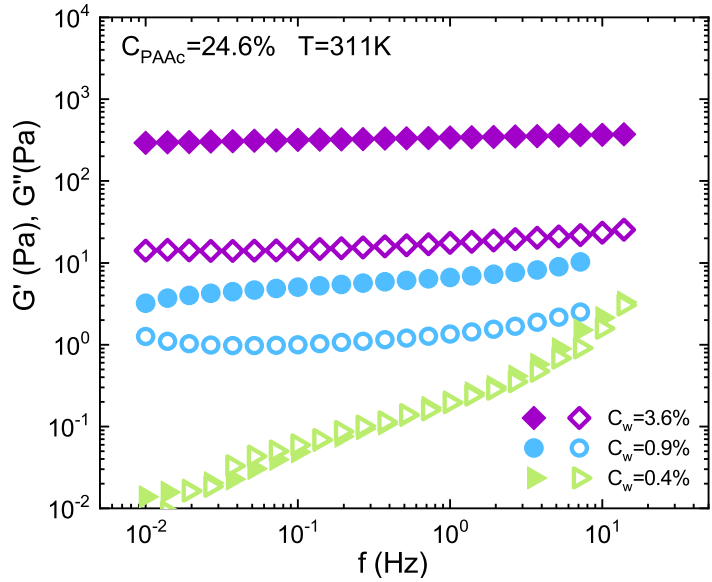}
\caption{Storage G$'$ (filled symbols) and Loss G$''$ (open symbols) moduli as a function of frequency for IPN microgels with C$_{PAAc}$=24.6$\%$ at concentration C$_w$=0.4$\%$ in the liquid,  C$_w$=1.0$\%$ in the glass and C$_w$=3.6$\%$ in the jammed state and at T=311$\unit{K}$ above the VPT.}
\label{fig7}
\end{figure}

Nonlinear oscillatory shear measurements in LAOS regime are shown in Figure \ref{fig8} where G$'$ and G$''$ are plotted as a function of shear strain at T=311$\unit{K}$ and at three different concentrations in the liquid, glass and jammed states.
Depending on microgel concentration, the magnitudes of G$'$ and G$''$ vary by several orders in agreement with Figure \ref{fig7}.
Moreover, the three curves show different features.
At low strains in the linear regime, the response at all concentrations is roughly a constant with G$''>$G$'$ at C$_{w}$=0.4$\%$ in the liquid state and G$'>$G$''$ at C$_{w}$=0.9$\%$ and C$_{w}$=3.6$\%$, in glassy and jammed state, respectively.
At larger amplitudes, for C$_{w}$=0.4$\%$ both G$'$ and G$''$ decrease as $\gamma$ increases, showing the so called strain thinning behaviour \cite{HyunJNNFM2002, KimRJ2002} typical of several fluid systems such as polymer melts, suspensions and solutions. Strain thinning, as well as shear thinning generally arises from polymer chain disentaglement and orientation with the flow direction. With increasing $\gamma$, the system flows more readily and consequently the moduli continue to decrease.
On the contrary, at C$_{w}$=0.9$\%$ and C$_{w}$=3.6$\%$ G$'$ decreases whereas G$''$ goes through a double peak at C$_{w}$=0.9$\%$ and a single peak at C$_{w}$=3.6$\%$ before decreasing definitely.
The arrows reported in Figure \ref{fig8} indicate the different breaking points related to the energy dissipation \cite{AppealSM2016, HelgesonJR2007, Derec2003} due to the induced deformation, that can be qualitatively viewed as a solid to fluid transition marked by a yielding behaviour \cite{Bonn2017, GhoshSM2019, MenutSM2012}.
The double peak at C$_{w}$=0.9$\%$ can be ascribed to attractive interactions among particles as also found in other systems \cite{Pham2008, AppealSM2016, UenoLang2010}. This behaviour can be explained considering that, by increasing $\gamma$, the first breaking point is due to the disruption of bonds between particles while the second one is associated to the breaking of the cage from which particles flow away. The system can be therefore considered an attractive glass.
Concerning higher concentrations, the single yielding displayed at C$_{w}$=3.6$\%$ is typical of systems characterized by repulsive interactions among particles \cite{Pham2008, VlassopoulosCOCIS2014} and it has been attributed to the breaking of the cage of the first neighbors surronding each particle. This peak, generally, results less pronounced as the attraction strenght among particles increases. However, since microgels above the VPT, as in our case, should behave as attractive spheres \cite{WuMacromol2003}, we cannot exclude that our system at this C$_{w}$ could be in a jammed state, due to both bonds and cage interactions, with a second peak expected, at higher strain, out of the experimental window in Figure \ref{fig8}.
This phenomenology is in agreement with the findings of Figure \ref{fig4} confirming that the system, with densely packed particles, is in the jammed state.
These findings further confirm the presence of two arrested states in IPN microgels with high PAAc content: an attractive glass for 0.6$\%<$ C$_{w} \leq$ 3.0$\%$ and a jammed state for C$_{w}>$ 3.0$\%$.

\section{Conclusions}\label{conclusions}

In conclusion, by investigating the rheological behaviour of interpenetrated polymer network microgels composed of poly(N-isopropylacrylamide) (PNIPAM) and polyacrylic acid (PAAc) at fixed PAAc content (C$_{PAAc}$) as a function of concentration in the range $0.05\%<$ C$_{w}$ $<7.2 \%$, we have found that at T=311$\unit{K}$, when particles are in the collapsed state, three different regimes occur: a non-Newtonian shear-thinning fluid for C$_{w} \leq 0.6 \%$ , an attractive glass characterized by a yield stress for $0.7\% \leq$ C$_{w}  \leq 3.0\%$, and a jamming state for C$_{w}>3.0 \%$. 
This phenomenology can be interpreted considering that in the dilute suspension regime, where 
interactions among particles are not favored, microgel suspensions have a purely viscous behaviour whose pseudoplastic nature can be associated to a shear-induced alignement of particles in the flow direction. By increasing concentration above the  glass transition, molecular interactions among microgels are enhanced and the exposure of PAAc dangling chains, resulting from the collapse of the thermosensitive PNIPAM, gives rise to the formation of an attractive glass. 
Finally, further compressing particles, they reach a mechanical contact undergoing a jamming transition. Both generalized glass ($\zeta_g$=1.56) and jamming ($\zeta_g \approx$ 7) volume fractions are well above 1 (and the typical values of hard spheres) as consequence of softness and deformability.
 In order to observe the same phenomenology at T=298$\unit{K}$, when particles are in the swollen state and are characterized by a higher degree of softness, much higher concentrations are required. 
It would be very intriguing to tune particles softness by changing temperature, pH and/or PAAc content to relate their rheological properties to particles softness.This could be the subject of  forthcoming studies.

\addcontentsline{toc}{chapter}{Bibliography}
\bibliography{bibliografia}
\bibliographystyle{unsrt}	

\end{document}